\documentstyle[preprint,prabib,aps]{revtex}
\begin{document}
\draft
\title{Electromagnetic field angular momentum in condensed matter
systems}

\author{Douglas Singleton \thanks{E-mail address :
dougs@csufresno.edu}}
\address{Dept. of Physics, CSU Fresno, 2345 East San Ramon Ave.,
Fresno, CA 93740-8031, USA}

\author{Jerzy Dryzek \thanks{E-mail address : JDRYZEK@alf.ifj.edu.pl}}
\address{Institute of Nuclear Physics, PL-31-342 Krakow, ul.Radzikowskiego
152, Poland}
\date{\today}
\maketitle

\begin{abstract}
Various electromagnetic systems can carry an angular momentum
in their {\bf E} and {\bf B} fields. The electromagnetic
field angular momentum (EMAM) of these systems can combine with
the spin angular momentum to give composite fermions or composite
bosons. In this paper we examine the possiblity that an EMAM
could provide an explanation of the fractional 
quantum Hall effect (FQHE) which is 
complimentary to the Chern-Simons explanation.
We also examine a toy model of a non-BCS superconductor
({\it e.g.} high $T_c$ superconductors) in terms of an EMAM.
The models presented give a common, simple
picture of these two systems in terms of an EMAM.
The presence of an EMAM in these systems might be
tested through the observation of the decay modes
of a charged, spin zero unstable particle inside
one of these systems.
\end{abstract}
\newpage

\section{Introduction}

The {\bf E} and {\bf B} fields of certain electromagnetic
configurations carry a field angular momentum.
This usually leads to interesting, but
not particularly physically relevant, results. For example,
it was noted long ago \cite{thom} that an electric
charge plus magnetic monopole
carried an EMAM in its fields which is independent
of the distance between the charges \cite{jack}. However, monopoles are
currently still only theoretical constructs, and therefore the physical
usefulness of this result is minimal. The more realistic
point charge/point magnetic dipole system also has an EMAM, which is
inversely proportional to the distance between the charge and
dipole \cite{sing}. On applying this charge/dipole result to
various atomic scaled systems it is found that classically
this EMAM tended to be small. For example, the EMAM arising from
the charge of the proton and the
intrinsic magnetic dipole moment of the electron in a hydrogen atom,
is on the order of $10 ^{-5} \hbar$  \cite{dry}.
The small size of this EMAM is due to the fact that
the charge/point dipole system field angular momentum
depends {\em inversely} on the distance between the charge and
dipole, and the typical atomic size
(0.1 nm) is ``large'' in this context.
One can also consider the EMAM of atomic scaled
systems arising from the interaction
between the charge of a nucleus
and magnetic dipole moments generated by the orbital
motion of the electrons, such as in hydrogen in an
excited ${\bf L} \ne 0$ state. The EMAM
of this last example is exactly zero \cite{dry}.
One place where a field angular momentum has
found an application is in the semi-classical approach
to diamagnetism \cite{zia}.

In nuclear scaled systems where
distances are five orders of magnitude smaller than for atomic
systems, there is a better chance that such an EMAM could
play a role \cite{sing}, and perhaps may have some connection to the
proton spin problem.

In this paper we wish to examine the possibility that
even though the EMAM apparently does not play a role
in individual atomic sized systems, it may nevertheless be
important in certain condensed matter systems. The idea
is that despite condensed matter systems
having distance scales which are of the 
atomic scale or larger, this can be countered by having
either an externally applied magnetic field, or by having
many contributions to the EMAM coming from the large
number of particles in a typical condensed matter system.
In effect the ``large'' distances are compensated for
by a large number of contributors to the EMAM. In the
following two sections we will apply these
ideas to the fractional quantum Hall system \cite{Tsui},
and in a toy model for non-BCS superconductors.

\section{FQHE in terms of EMAM}

In the Chern-Simons models of the FQHE
composite objects are formed which consist
of electrons bound to fictitious Chern-Simons flux tubes \cite{gir}
\cite{kane}. When these composite fermions are placed in an
external magnetic field there is a partial cancellation or
screening of the Chern-Simons flux by the external magnetic
flux. For certain values of the external magnetic field this
leads to these composites of a charge $+$ Chern-Simons
flux tube undergoing the integer quantum Hall effect with
respect to the uncanceled or unscreened part of the
external magnetic field. These explanations have their
theoretical origins in work by Wilczek \cite{wilc1} where
the electric charge sits {\em outside} the region of magnetic
flux. In this paper we take the electron as sitting {\it inside}
the magnetic flux tube, so that the system develops
an EMAM which combines with the spin of the electron
to give rise to a composite fermion which then undergoes
the IQHE. This EMAM explanation of the FQHE is meant
to be complimentary to the Chern-Simons explanation of
the FQHE (Ref. \cite{shs} gives an overview of
the aspects Chern-Simons theory of the FQHE which
will be relevant in this paper).
The possible advantage of the EMAM
explanation is that we do not need to postulate a
fictitious Chern-Simons flux -- the composite fermion
arises directly from the combination of the EMAM plus the
spin of the electron.

In this paper we will work mainly in the mean field
approximation where the interaction between
electrons is ignored.
In this approximation the Chern-Simons approach to
the FQHE starts with a Hamiltonian of the form
\begin{equation}
\label{cherns}
H_{mf} = \sum_i {[{\bf p}_i +\frac{e}{c} {\bf A} ({\bf r}_i)
-\frac{e}{c} {\bf a} ({\bf r}_i)]^2 \over 2m_b}
\end{equation}
where ${\bf r}_i$ is the position of the $i^{th}$ electron;
${\bf A} ({\bf r}_i)$ is the external vector potential associated
with the external magnetic field, and ${\bf a} ({\bf r}_i)$ is
the Chern-Simons vector potential; $m_b$ is the effective mass
of the electron in the material. Usually there is an
electron-electron interaction term of the form
$\sum_{i<j} V({\bf r}_i - {\bf r}_j)$, with $V$ being
a Coulomb potential, for example. In the mean field approximation
this term is dropped. The original Hamiltonian for the
system is of the form given in Eq. (\ref{cherns}) except
with no ${\bf a} ({\bf r}_i)$ term. This term comes into
the Hamiltonian in the following way.
If $\Psi ({\bf r_1 , r_2 ,
... , r_N}) $ is a solution to the original Hamiltonian
($H_o \Psi = E \Psi$) with no ${\bf a} ({\bf r}_i)$,
then the phase transformed wavefunction
\begin{equation}
\label{phaset}
\Phi ({\bf r_1 , r_2 , ..., r_N}) =
\left[ {\prod _{i<j}}  e^{-i2m \theta ({\bf r}_i - {\bf r}_j)} \right]
\Psi ({\bf r_1 , r_2 , ... , r_N})
\end{equation}
will be a solution to the Hamiltonian in Eq. (\ref{cherns}).
In Eq. (\ref{phaset}) $m$ is an integer and $2m$ is the
(even) number of Chern-Simons flux tubes attached to the
electron. $\theta ({\bf r}_i - {\bf r}_j)$ gives the angle
between the vector ${\bf r_i - r_j}$ and the ${\bf {\hat x}}$
axis and is defined modulo $2 \pi$ (${\bf {\hat z}}$ is
taken as perpendicular to the 2D system and ${\bf {\hat x}}$
is in the plane). This transformation can be thought of as a
singular gauge transformation since $\theta$ is nonsinglevalued.
The Chern-Simons vector potential is then given by
\begin{equation}
\label{cspoten}
{\bf a} ({\bf r}_i) = i \nabla _i \left[ {\prod _j}  e^{-i2m \theta ({\bf r}_i)}
\right]
\end{equation}
Since this vector potential is a gradient one would
think that the ``magnetic'' field associated with it is
zero, ${\bf b_{CS} (r)} = \nabla \times {\bf a (r)} =0$. However,
due to the singular character of the transformation
in Eq. (\ref{phaset}) one finds that ${\bf b_{CS} (r)} =
2 \pi (2 m) n({\bf r})$, where $n ({\bf r})$ is the local
electron density which is a sum of delta function that
spike  at the location of each electron. In the mean
field approximation $n ({\bf r})$ is replaced by
$n_e \approx const.$, the average density.

In this Chern-Simons mean field approach the FQHE
is described  as an integer quantum Hall effect
(IQHE) for the composite
fermions of the electrons $+$ Chern-Simons flux
tubes. When the external magnetic field is imposed
it partially cancels or screens the Chern-Simons field
so that the composite fermions see a reduced, effective
magnetic field of $B_{eff} = B_{ext} - b_{CS}$ with
respect to which these composite fermions then undergo the
IQHE. Although this mean field description is very crude
(since it ignores the inter-electron interactions) it
nevertheless gives the Jain series \cite{jain} for
the fractional quantum Hall states, and it
provides a starting point for more accurate studies
such as using the random phase approximation (RPS) \cite{shs}.

We now want to present a similar picture of the FQHE,
but without the need of introducing the fictitious Chern-Simons
potential. In our approach the composite fermions
arise through the combining of the intrinsic
spin of the electron with the EMAM produced
by the external magnetic flux and the charge
of the electron which sits {\it inside} a
magnetic flux tube. That such a complimentary
view is possible is suggested by
Ref. \cite{wilc1} where the charge $+$ flux tube
system is equivalently examined using both
singular gauge transformation arguments (Chern-Simons
type description) {\it and} angular momentum arguments (EMAM
type description).

The angular momentum carried in the electric and
magnetic fields can be written as \cite{jack}
\begin{equation}
\label{1}
{\bf L}_{em} = {1 \over 4 \pi c} \int {\bf r \times (E \times B)}
d^3 {\bf r}
\end{equation}
We consider an electron located in the $z = 0$ plane
whose electric field in cylindrical coordinates takes
the usual Coulomb form
\begin{equation}
\label{2}
{\bf E} = {- e \over r^3} (\rho {\hat {\bf \rho}} + z
{\hat {\bf z}} )
\end{equation}
The external magnetic field, {\bf B},is taken to be of uniform
strength, $B_0$, inside an infinitely long cylinder of radius $R$,
and zero outside this cylinder. The {\bf B} field
points along the positive $z$ axis, and the axis of the
cylinder goes through the electron. In cylindrical
coordinates
\begin{equation}
{\bf B}=
\left\{
\begin{array}{ll}
B_0 {\hat {\bf z}} &  \; \; \;  r\le R \\
0  &  \; \; \;  r > R
\end{array}
 \right.
\label{3}
\end{equation}
>From Eqs. (\ref{2}) and (\ref{3}) ${\bf E} \times {\bf B} \propto
{\bf {\hat \phi}}$. When this is combined with ${\bf r} = \rho
{\bf {\hat \rho}} + z {\bf {\hat z}}$ in ${\bf r \times (E \times B)}$
one gets components in the ${\bf {\hat z}}$ and ${\bf {\hat \rho}}$
directions. The ${\bf {\hat \rho}}$ component will vanish upon
doing the $\phi$ integration in Eq. (\ref{1}). The remaining
${\bf {\hat z}}$ term gives on combining Eqs. (\ref{1}) - (\ref{3})
\begin{equation}
\label{4}
L^{em}_z = { e B_0 R^2 \over 2 c} = { e \Phi \over 2 \pi c}
\end{equation}
where $\Phi$, is the flux
of the magnetic tube. From Eq. (\ref{3})
the direction of this field angular momentum is
determined by the direction of ${\bf B}$ and the
sign of the charge. If $\Phi$ equals some integer
multiple, $n$, of the quanta of
magnetic flux ($ \Phi_0= 2 \pi \hbar c/e$)
then $L^{em}_z= n \hbar$.
The spin of the electron is $S_z=\pm\hbar/2$ so the
combination of the EMAM plus the electron's spin
yields a composite fermion with a half integer
angular momentum of $(n \pm \frac{1}{2}) \hbar$.
Thus for certain values the
external magnetic flux the charge of the
electron will ``absorb'' an integer
multiple of the magnetic flux quanta, $\Phi _0$,
and generate an EMAM. This EMAM, when combined with the
intrinsic spin of the electron, gives a composite fermion,
which then undergoes the IQHE with respect to the
remaining, ``unabsorbed'' flux, $\Phi_{ext} - n \Phi_0$.
The phase factor of the wave function of this composite
fermion equals (-1)$^{1+2n}$, which explains the
Jain series  \cite{jain} in a simple way. Also
according to Laughlin \cite{Lau} the excited-state
of a 2D electron gas in a strong magnetic field
has fractional charge. Thus if we replace $e$ with
$\frac{1}{3}e$ in (\ref{4}) we obtain 
$L^{em}_z =\frac{1}{3}\hbar$ when $\Phi=\Phi_0$. This yields
fractional statistics for the excited-state of the 2D
electron gas in a strong magnetic field, which has
been observed experimentally \cite{Chang}.

Eq. (\ref{4}) is for a free electron,
but we can extend our considerations slightly
beyond the mean field approximation by including
screening effects. For an
electron inside some material the electric field will
be screened past a certain distance as follows:
\begin{equation}
\label{6}
{\bf E} = -{e \over \epsilon_s r^2} (1 + \lambda r) e^{-\lambda r}
{\bf {\hat r}},
\end{equation}
where $\lambda^{-1}$ is  the screening length and $\epsilon_s$ is the
dielectric constant. There are two cases connected with this
effective screening distance. First if the screening length
is much larger than the radius of the magnetic flux tube
($\lambda^{-1} \gg R$) then Eq. (\ref{4}) and all the developments
leading up to it should still hold approximately. The second case is when
the screening length is much shorter than  the radius of the
magnetic flux tube ($\lambda^{-1}\ll R$). In this case one can repeat
all the developments up to Eq. (\ref{4}), but now one finds that
\begin{equation}
\label{7}
L^{em}_z = {e B_0 \over 2 c} \left({2 \over \lambda \sqrt{\epsilon_s}} \right)^2
\equiv {e B_0 R_{eff} ^2 \over 2 c} = {e \Phi _{eff} \over 2 \pi c}
\end{equation}
This result is similar to that of Eq. (\ref{4}) except
with the replacement $R \rightarrow R_{eff}
= 2/\lambda \sqrt{ \epsilon_s}$, and $\Phi \rightarrow
\Phi _{eff}$. The difference is that in Eq. (\ref{4})
the radius is set by the arbitrary, external choice of
$R$, while in Eq. (\ref{7}) the effective radius, $R_{eff}$,
is set by the screening length of the material.

As in the previous case we want $L^{em} _z = n \hbar$, so
one can ask if Eq. (\ref{7}) can be satisfied for
reasonable values of the magnitudes of $B_0$ and
$R_{eff} ={2/ \lambda \sqrt{\epsilon_s}}$.
As an example take $n=1$ so that $L^{em} _z = \hbar$. Eq.
(\ref{7}) then implies that we want $e B_0 R_{eff} ^2
/ 2 c = \hbar$. For the FQHE one has magnetic fields
on the order of $B_0 \simeq 100$ kilogauss or larger.
The FQHE is associated with semiconductor materials
such as silicon or gallium-arsenic. For silicon the
screening length is on the order of $\lambda ^{-1} =$
24 nm and $\epsilon_s=12$ which implies 
$R_{eff} = {{2}/{ \lambda \sqrt{\epsilon_s}}}
=1.39 \times 10^{-6}$ cm.
Using these numbers in Eq. (\ref{7}) gives an EMAM of
$1.5 \times 10 ^{-27}$ erg-sec which is of order $\hbar$
as desired.

The EMAM explanation of the FQHE
is a simple, alternative
way of looking at this effect, which is
complimentary to the Chern-Simons approach in
the mean field approximation.
A strong point of the EMAM explanation
is that it does not require one to postulate
a Chern-Simons flux; one only needs the real
external magnetic flux and the internal spin of the
electron.

\section{Superconductors in terms of EMAM}

Ordinary superconductors such as Mercury or
Tin are well explained by BCS theory. However,
with the discovery of high temperature superconductors,
and heavy-fermion superconductors \cite{physrep}
one finds systems for which it is no longer certain
that BCS theory is the full explanation. In particular
it is not clear how the BCS model can lead to
the binding of Cooper pairs at the large
temperatures found in high $T_c$ materials.
One possibility is that the
phonons in high $T_c$ materials have some strong self
coupling similar to that of gluons in QCD, leading
to a much stronger binding \cite{vlad}. 
In this section we examine a toy model of a 1D and 2D atomic
lattice. The atom at each lattice site is assumed to possess
a magnetic dipole moment with a magnitude of the order of
the Bohr magneton, $\mu _B = e \hbar / 2 m_e c$.
We then consider an electron placed in this
lattice of atomic magnetic dipoles. The charge, $-e$, of the
electron in conjunction with the magnetic dipoles at each
lattice site will generate an EMAM. If the sum total EMAM
is of the right size then the combination of electron
spin plus total EMAM will result in an integer
angular momentum making the electron effectively a
boson, which can then condense into the superconducting
state. Heuristically one can think of this as the
electron ``binding'' with ({\it i.e.} sitting in)
some local region of the lattice, leading to a bosonic
composite with an integer angular momentum.

We first examine the EMAM carried in the field of an
electron with charge, $-e$, and a {\it single}
point magnetic dipole, ${\bf m}$.
Without loss of generality we place the magnetic
dipole at the origin and align its magnetic moment along
the z-axis. Its magnetic field takes the standard form
${\bf B} = {1 \over r^3} [3 ({\bf m \cdot {\hat r}}) {\bf {\hat r}}
- {\bf m}]$. Placing the electron at some arbitrary point
${\bf R}$ gives an electric field of the form
${\bf E} = -e {{\bf {\hat r}'} \over {r'}^2}$ where ${\bf r' =
r - R}$. Using these two fields
in Eq. (\ref{1}) yields an EMAM 
whose z-component is \cite{sing}
\begin{equation}
\label{emam}
(L_{em})_z = \pm {e \vert {\bf m} \vert \over c R} (1 - cos ^2 \theta ) 
\end{equation}
$\theta$ is the angle between ${\bf m}$ and ${\bf R}$.
The $\pm$ sign indicates that the direction of
$(L_{em})_z$ depends on the direction of ${\bf m}$
and on the sign of the charge.
We now want to apply Eq. (\ref{emam}) to an
electron sitting in a 1D antiferromagnetic lattice of 
atomic magnetic moments. We will take the magnetic moments
to alternate between ${\bf m}_a = a \mu _B {\hat {\bf z}}$
and ${\bf m}_b = -b \mu _B {\hat {\bf z}}$ on
neighboring lattice sites; $\mu _B$ is
the Bohr magneton; $a, b$ are positive constants of order unity;
${\hat {\bf z}}$ is perpendicular to the line of the 1D
lattice. The lattice spacing between neighboring
${\bf m}_a$'s and ${\bf m}_b$'s is $d$.
Since the magnetic moments are taken to be aligned along
the z-axis, $\theta = \pi /2$ in Eq. (\ref{emam}). The
EMAM due to the $n^{th}$ plus $(n+1)^{th}$ lattice site is
\begin{equation}
\label{emam1}
(L_{em})_{zn} =  {e \vert {\bf m}_a \vert \over c R_n}
- {e \vert {\bf m}_b \vert \over c R_{n+1}}=
{r_o \over d } \left( {a \over n} - {b \over  (n+1)} \right)
{\hbar \over 2}
\end{equation}
where $r_o = e^2 /m_e c^2$ is the classical electron radius;
${\bf m}_a$ is located at the $n^{th}$ lattice site while
${\bf m}_b$ is located at the $(n+1)^{th}$ lattice site.
There will also be an EMAM from the
interaction between the electron and the magnetic moment
of the lattice site on which it is located. From Eq. (\ref{emam})
it may appear as if this would lead to a divergent EMAM.
If the electron and the magnetic dipole are placed on the
same lattice site then apparently $R \rightarrow 0$ in
Eq. (\ref{emam}) implying that $(L_{em})_z \rightarrow \infty$.
This assumes that the point dipole expression for
the magnetic field ({\it i.e.} ${\bf B} = {1 \over r^3} [3
({\bf m \cdot {\hat r}}) {\bf {\hat r}} - {\bf m}]$) is
valid for all $r$. However, once $r$ takes values which are
``inside'' the electronic orbitals of the atomic magnetic
dipole in question this expression for ${\bf B}$ is no longer
valid. If one models these electronic orbitals as effective
currents then as the external electron moves closer to the
origin of the atom ($r=0$) then the EMAM coming from the
``inside'' magnetic field tends to {\it cancel} the
EMAM coming from the ``outside'' magnetic field. For a multi-electron
atom this is difficult to see in detail, however in
ref. \cite{dry} it was shown that for the hydrogen atom
the EMAM arising from the magnetic field of an electron
in a non-zero orbital angular momentum state and the electric
field arising from the charge of the proton was exactly
zero. This was due to a cancellation between the different
EMAMs that resulted from the different magnetic fields
in the regions ``inside'' and ``outside'' of the electron's
orbit. Generally as the external
electron is moved closer to the atomic magnetic dipole
the EMAM will tend to increase until the electron is at a distance
which is just outside of the outer electronic orbitals of the
atom. As the electron moves inside the electronic orbitals the
net EMAM will decrease due to the cancellation between
the EMAMs from the different magnetic fields inside and
outside the electronic orbitals. In dealing with the
EMAM from the electron and the atomic magnetic moment
which are at the same lattice site we will assume
that the dipole moment is ${\bf m}_b$, and that the
electron is at a distance, $d_0$, which is roughly
the atomic radius of the magnetic moment. This will make
the contribution from this lattice site as large as
possible. However, for this 1D case we will find that
whether or not one maximizes the EMAM from this closest
lattice site, that the net field angular momentum is still too
small to play a significant role. Putting this all together
the net EMAM for this 1D lattice is
\begin{equation}
\label{emam1D}
L_{em}^{(1D)} = - {r_o b \over d_0} \left({\hbar \over 2} \right)
+ {r_o \hbar  \over d} \sum_{n=1}^{\infty}
\left({a \over 2n-1} - {b \over 2n} \right)
\end{equation}
The first term is the maximum possible contribution to
the EMAM arising from the interaction between the electron
and the lattice site on which it sits. 
A factor of two in the second term comes from summing the
lattice sites to the right and left. The sum in
Eq. (\ref{emam1D}) is divergent unless $a=b$.
We will assume that $a \ne b$. In physical
situations this sum is cut off from above for several
reasons : the screening of the electric field
of the electron, the finite size of the sample, or
the appearance of defects
 of the crystalline lattice
 which would alter the ideal
antiferromagnetic ordering which we have assumed.
Depending on this cut-off length, $L_{em}$ could take on
any value up to $\infty$. As a specific example assume
$a=4$ , and $b=2$.
Take the lattice spacing as $d=2 \times 10 ^{-10} m$,
take $d_0 \simeq 0.53 \times 10^{-10}$ m (the Bohr
radius), and $r_o \simeq 2.82 \times 10 ^{-15} m$. Finally we
assume that the sum in Eq. (\ref{emam1D}) is cut off
after $10^{20}$ magnetic dipole lattice sites.
With this the sums in Eq. (\ref{emam1D}) can be approximated
by $(a-b) \int _1 ^{10^{20}} {dx \over 2 x} \simeq 46.1$.
With these numbers $L_{em}\simeq 0.001 {\hbar \over 2}$
which is small compared to the fundamental unit of angular
momentum (${\hbar \over 2}$). The cut-off of
$10^{20}$ lattice sites is arbitrary, however, because of the
logarithmic behavior of the integral approximation for the sum
in Eq. (\ref{emam1D}) even taking the cut-off at
$10^{100}$ lattice sites would not make $L_{em}$ of the
order $\hbar$. Thus given this small magnitude
for $L_{em}$ it is unlikely that the EMAM plays any
significant role in this 1D case.

Next we consider a 2D lattice in which all the
magnetic moments are aligned in the same direction.
This is somewhat similar to the 2D Ising
model of magnetic moments on a lattice. For this
2D case one would expect $L_{em}$ to become larger since there
are more neighboring dipoles for the electron to interact
with. We take the dipoles with ${\bf m}_a = a \mu _B {\hat {\bf z}}$
and $a=4$ as before. Under these conditions the
EMAM along the z-axis from the dipole located at $x = nd$ and $y =md$ is
\begin{equation}
\label{emx}
(L_{em}^z )_{n,m} = {e \vert {\bf m} \vert \over \sqrt{(nd)^2 + (md)^2}} =
{r_o a \over d \sqrt{n^2 + m^2}} {\hbar \over 2}
\end{equation}
Summing the first $N$ sites along the $x$ and $y$ directions
gives a total EMAM of
\begin{equation}
\label{emam2da}
L_{em} = \sum_{n, m=-N}^{N} (L_{em}^z)_{n,m}  = 
{2 r_0 a \over d} \left({\hbar \over 2} \right)
\sum_{n, m=1}^{N}
{1 \over \sqrt{n^2 + m^2}}
\end{equation}
There should also be a contribution
coming from the interaction between the electron and
the dipole located on the lattice site at which
the electron is located, exactly as in the 1D lattice case
({\it i.e.} one should add a term to Eq. (\ref{emam2da})
like the first term in Eq. (\ref{emam1D})). This term,
by itself, gives only a small contribution, and we
will therefore ignore it. The double sum in Eq. (\ref{emam2da})
can be approximated by the double integral $\int_1 ^N \int_1 ^N
[(dx) (dy) / \sqrt{x^2 +y^2}] \simeq \int _0 ^{2 \pi} (d \phi)
\int _1^N (r dr) {1 \over r} = 2 \pi (N-1)$ where
$r = \sqrt{x^2 +y^2}$. In replacing the Cartesian integral
over $dx dy$ by the polar integral over $r d \phi dr$
we are ignoring the contributions from the dipoles at the
corners of the square region indicated by the sum in
Eq. (\ref{emam2da}). This is actually physically more
realistic, since the dipoles which influence the
charge should have a circular, not a square, cut-off.
For an infinite 2D lattice ({\it i.e.} $N \rightarrow
\infty$) the field angular momentum will diverge (this time
linearly rather than logarithmically) unless the integral is
cut-off from above. Physically this cut-off will arise
due to the screening of the electric field of the
electron. Here we will simply put in this screening
distance cut-off by hand to see roughly the
size needed to generate an EMAM of the order
$\hbar/2$. Setting the cut-off in the above
integral approximation of the sum as $N = 1500$, and taking
the other parameters as in the 1D example, gives
an EMAM of $L_{em} \simeq 1.06 (\hbar /2)$.
Multiplying this $N$ with the lattice
spacing of $d = 2 \times 10^{-10}$ m implies
a screening distance of $3 \times 10^{-7}$ m. The combination
of this EMAM plus the internal spin of the electron yields
a composite with integer spin. This bosonic quasi-particle
of electron plus the local region of the magnetic dipole
lattice in which it sits, could condense into a
superconducting state. Although the screening distance
cut-off in the above example (which was set by hand)
was large, it was not so large as to be out of the question
for some materials. For example, it is only somewhat larger
than the screening length of silicon discussed at the end
of section II (this is only for comparison since
silicon is not a high temperature superconductor).
This does imply that metallic materials or materials
with small screening lengths would not be viable under
the above mechanism.

One may wonder if a ferromagnetic 3D lattice might not work
even better at giving a substantial EMAM. There are two arguments
as to why this proposed mechanism would not work
better for a 3D ferromagnetic lattice. First, the EMAM of the
magnetic dipole and electric charge is maximized when
$\theta = \pi /2$ as occurs when the electron is in the same
plane as the dipoles {\em and} if the dipoles have their moments
oriented perpendicular to this plane. For a 3D lattice of
magnetic dipoles, the dipoles which are not coplanar with
the electron will give a smaller EMAM. In addition the EMAM
from the magnetic dipoles above the plane of the electron will
tend to cancel against the EMAM from the magnetic
dipoles below the plane of the electron.
Second, the size of the EMAM depends closely on $N$ from
Eq. (\ref{emam2da}) which in turn is connected with the
screening distance of the electric field in the material.
For ferromagnetic materials, such as iron, these screening
distances are small and so $N$ would also in general be
too small to generate a large EMAM.

>From the above estimates we can
see that for a 1D lattice it is unlikely
that the EMAM could play any significant
role since the number of dipole lattice
sites ($N$) which must contribute is just too large.
For a 2D lattice the number of lattice sites
needed to give an EMAM of order $\hbar /2$ is
much smaller (since the EMAM dependence goes
from $ln (N) \rightarrow N$ on going from
$1D \rightarrow 2D$). However, $N$ (or the screening
length) must be ``fine tuned'' in order to get
an EMAM of exactly some integer number times
$\hbar /2$. Also the high $T_c$ materials to which
we want to apply this mechanism are generally
antiferromagnetic rather than ferromagnetic
as in the simple 2D case we considered. To address
this one could group pairs of anti-aligned lattice sites.
The closer lattice site would contribute more than
the further lattice site due to the $1/r$ dependence
of the EMAM in Eq. (\ref{emam})
so that each pair would give some net EMAM, just as
in the 1D antiferromagnetic case. Each
successive pair of sites would also contribute
an EMAM in the same direction as the closest
pair. For such an anti-ferromagnetic 2D lattice one
would have to increase the number of lattice sites
so that the total EMAM would have the right
magnitude to give a net angular momentum which had
integer (bosonic) values. Thus the case for this alternative
mechanism of generating a superconducting
state may appear only marginally viable. Nevertheless,
this idea does have two interesting features :

\begin{enumerate}

\item
There is no need to postulate any kind
of exotic binding mechanism for Cooper pairs
which persists up to the temperatures
observed in high $T_c$ materials. As long as the electron
is ``bound'' to ({\it i.e.} sits in) an {\em ordered} lattice,
as discussed above, then it will effectively act as
a boson which can condense into a superconducting state.
In this picture the reason for the disruption
of the superconducting state is the disordering of the
lattice and/or a changing of the screening length
as the temperature increases.
The increase of the temperature can also influence
$L_{em}$ through changes of $R$ and $\theta$ in Eq.
(\ref{emam})

\item
It provides a common picture for high $T_c$
materials and the fractional quantum Hall systems.
In the arguments given above both systems
are explained in terms of the combination of
EMAM with intrinsic spin angular momentum to
transform fermions into composite fermions or
composite bosons.

\end{enumerate}

One final problem in using this EMAM model
for high $T_c$ materials is that
experiments indicate that the charge of the
order parameter in high $T_c$ materials is still
$-2e$, which is an indication of Cooper pairing. In the EMAM
picture, as in the holon/spinon picture \cite{laugh},
the order parameter charge equals the
charge of the electron, $-e$. One possible
resolution to this is that in certain situations
the electric charge of an object may be shifted
through the effects of the background or vacuum. An example of
this is the $\Theta$ vacuum effect \cite{witten}
where a magnetic charge placed in a vacuum with a
non-zero, CP violating parameter ($\Theta$) will
have its electric charge shifted by $-e \Theta / 2 \pi$.
The present situation is different from the $\Theta$
vacuum example, since in that case it is
a hypothetical monopole which develops a shifted electric
charge. Nevertheless, in the present context one
could postulate that if the high $T_c$ materials
behaved as an ``exotic'' background that the effective
charges of the electrons could be shifted via
their interaction with this background. In this
regard it should be mentioned that recently it has
been observed that high $T_c$ materials do violate
$P$ and $T$ invariance \cite{ptv}.

\section{Possible experimental test for the EMAM
mechanism}

An experimental test, which in principle would probe
the existence of an EMAM is to observe the decay of a low speed,
unstable, charged  and
zero spin particle, such as a pion ($\pi ^{\pm}$),
inside the magnetic field environment
of one of these systems. Considering the example of
the positively charged pion, in vacuum
it decays into an antimuon and a muon neutrino.
These decay products move in opposite directions,
with anti-aligned spins in order to conserve
momentum and angular momentum. If the pion is
placed inside a quantum Hall system at a
quantum Hall step, or in a strong magnetic field,
then it will behave as an effective boson, but now
with a non-zero spin coming from the EMAM.
If the magnetic field is such that the EMAM associated with
the pion is $\hbar$ then the effective spin of the pion
will be spin $1 =$ spin $0 +$ EMAM. The decay of this
new ``spin 1'' pion will be altered  from the decay of the
pion in free space. In free space when the pion decays the
positively charged antimuon, $\mu ^+$, and the muon neutrino,
$\nu _{\mu}$, move off back to back {\it in any direction}.
and have their spins anti-aligned. In the case when the
pion is implanted into a quantum
Hall system, an EMAM develops, and the direction in which
the decay products move off becomes restricted. In the above
example the initial angular momentum of the system
was $\hbar$ and pointed along the direction of the
magnetic field. Thus the decay products also had their
angular momentum (EMAM plus the spins of both the antimuon and
muon neutrino) aligned along or against the direction of the magnetic
field. Neutrinos in the Standard Model are purely left handed,
which means that their spin and their momentum
point in opposite directions (If neutrinos turn out to have
a small mass, as recent experiments indicate, then this is only
approximately true. However, in the case of the pion
decay the neutrino is very well approximated as being purely
left handed). Since the spin of the neutrino
is required to be along or against the direction of the magnetic
field this means that its momentum must also lie along the
line of the magnetic field. Thus in the presence of an EMAM the
antimuon and muon neutrino should decay predominately with their
momenta along the direction of the magnetic field. If there
is no EMAM then the two decay products should be able to move
off in any direction since there is no angular momentum
quantization axis picked out before hand by virtue of the
fact that the pion is spin zero.

\section{Discussion and Conclusions}

In this article we have given an alternative, simple
picture of the fractional quantum Hall effect. We looked
at the EMAM which arises when an electron
{\it sits inside} a magnetic flux tube. The size of
the EMAM depends on the amount of flux, $\Phi$,
in which the electron is embedded. For certain
values of the flux the EMAM
takes on integer multiples of $\hbar$. This EMAM
then combines with the intrinsic spin of the
electron to yield a composite fermion. The FQHE occurs when
some number of electrons sits inside
a flux tube of the appropriate strength
to give rise to an EMAM, which when added
to the internal spin of the electrons
gives the combined system a net half integer
spin making it effectively a composite fermion.

Next a toy model for non-BCS superconductivity was
given in terms of an EMAM. By considering an
electron sitting in a 1D or 2D lattice with
anti-aligned/aligned atomic, magnetic dipoles
at each site it was found that an EMAM
occurred which, when combined with the
internal spin of the electron, could yield
an integer angular momentum. The
combination of the electron plus the
``local'' region of the lattice in which
it was located, acted effectively as an integer spin,
bosonic quasi-particle, which could condense into
a superconducting state. The main weakness of this
argument, even for the 2D lattice, was the large number of
lattice sites ($\simeq N^2 \approx 10^6$) which
needed to be included in the ``local'' neighborhood of the
electron in order to get an EMAM of the
correct order to transform the electron
into a boson. The chief advantage of the EMAM mechanism for
high $T_c$ superconductivity is that it does not
require some exotic binding mechanism to form Cooper pairs,
but rather it is connected with the magnetic ordering of the lattice.

The presence of an EMAM might be tested by the observation
of the decay of unstable, charged, zero spin particles,
such as pions, within a quantum Hall system
or a strong magnetic field.
In such an environment the direction of the decay products
of the unstable particles would tend to be restricted by the direction
of the magnetic field, which determines the direction of the EMAM.

\section{Acknowledgment} It is pleasure to thank K. Ruebenbauer for useful
discussions in connection with certain aspects of this work.


\newpage

\begin{thebibliography}{AL}

\bibitem{thom} J.J. Thomson, {\it Elements of the Mathematical Theory
of Electricity and Magnetism} (Cambridge U.P.,Cambridge, 1904), 3$^{rd}$
Ed., Sec. 284

\bibitem{jack} J.D. Jackson, {\it Classical Electrodynamics} (Wiley,
New York, 1975), 2$^{nd}$ Ed., p. 251

\bibitem{sing} D. Singleton, Phys. Lett. {\bf B427}, 155, (1998);
Am. J. Phys. {\bf 66}, (1998)

\bibitem{dry} J. Dryzek and D. Singleton, Am. J. Phys., {\bf 67}, 930
(1999)

\bibitem{zia} S.L. O'Dell and R.K.P. Zia, Am. J. Phys., {\bf 54},
32 (1986)

\bibitem{Tsui} D.C. Tsui, H.L. Stormer and A.C. Gossard, Phys. Rev.
Lett. {\bf 48} 1559 (1982).

\bibitem{gir} S.M. Girvin and A.H. MacDonald, Phys. Rev. Lett., {\bf 58},
1252 (1987)

\bibitem{kane} D.H. Lee and C.L. Lane, Phys. Rev. Lett., {\bf 64},
1313 (1990)

\bibitem{wilc1} F. Wilczek, Phys. Rev. Lett., {\bf 48}, 1144 (1982);
Phys. Rev. Lett., {\bf 49}, 957 (1982)

\bibitem{shs} Steven H. Simon, cond-mat/9812186

\bibitem{jain} J.K. Jain, Phys. Rev. Lett. {\bf 63}, 199 (1989)

\bibitem{Lau} R.B. Laughlin, Phys. Rev. Lett. {\bf 50}, 1395 (1983)

\bibitem{Chang} A.M. Chang, M.A. Paalanen, H.L. Str\"ormer,
J.C.M. Hwang, D.C. Tsui, Surface Science, {\bf 142}, 173 (1984)

\bibitem{physrep} G.R. Stewart, Rev. Mod. Phys., {\bf 56}, 755 (1984)

\bibitem{vlad} V.D. Dzhunushaliev, Phys. Rev. {\bf B 54}, 10121 (1996)

\bibitem{laugh} R.B. Laughlin, Science, {\bf 242}, 525 (1988)

\bibitem{witten} E. Witten, Phys. Lett. {\bf B86}, 283 (1979)

\bibitem{ptv} Y. Maeno, {\it et. al.}, Nature {\bf 372}, 532 (1994);
G.M. Luke, {\it et. al.}, Nature {\bf 394}, 558 (1998);
K. Krishana, {\it et. al.}, Science, {\bf 277}, 83 (1997)

\end{thebibliography}
\end{document}